\documentstyle[12pt]{article}
\begin{document}
\hfuzz=1pt
\setlength{\textheight}{8.5in}
\setlength{\topmargin}{0in}
\begin{centering}
\LARGE {\bf Quantum cryptography based on Wheeler's delayed choice
experiment} \\  \vspace{.75in}
\Large M. Ardehali \\ \vspace{.5in}
\large Microelectronics Research Laboratories,
NEC Corporation,
1120 Shimokuzawa,
Sagamihara, \\
Kanagawa 229
Japan \\ \vspace{.75in}
\end{centering}

\begin{abstract}
We describe a cryptographic protocol in which
Wheeler's delayed choice experiment is used to generate
the key distribution. The protocol, which uses photons polarized
only along one axis, is secure against general attacks.
\end{abstract}

\pagebreak

In 1970, Wiesner wrote a highly innovative paper about quantum
cryptography, introducing a new branch of Physics
and computation.
Unfortunately, his idea went unnoticed, and his paper
was not published until 1983 \cite{1}).
Wiesner's idea was brought back to
life in 1980s primarily by the work of
Bennett, Brassard \cite{2}.
Bennett {\em et al.} have recently
reported the experimental
realization of the first quantum cryptographic protocol
\cite{3}.

Theoretical models for quantum key distributions have been proposed
based on uncertainty principle \cite{2}, EPR \cite{4} states \cite{5},
and any set of two non-orthogonal states \cite{6}. Here we
describe a model for quantum key distribution
based on Wheeler's delayed
choice experiment.

Before proceeding,
let us briefly describe Wheeler's delayed choice experiment (for
detailed explanation see \cite{7}, especially Fig. $4$; for
consistency, we use
Wheeler's notations).
In the first arrangement,
a single photon (or low intensity light pulse) comes in at $1$
and encounters a beam splitter $\frac{1}{2}S$ which splits it
into two beams, $2a$ and $2b$, of equal intensity (See
Fig. $4$ of \cite{7}). The beams are
reflected by mirrors $A$ and $B$ to a crossing point and are then
detected by detectors $1$ and $2$.
Thus one finds out by which route ($2a$ or
$2b$) the photon came.
In the second arrangement, a beam splitter $\frac{1}{2}S'$
is inserted at the point
of crossing in front of detectors.
The beams $2a$ and $2b$ are brought into constructive
interference so that a count is always triggered
from detector $1$. Thus in
this arrangement, one concludes that the photon came by both
routes. In Wheeler's experiment:
\begin {quote}
In the new ``delayed-choice'' version of the experiment one decides
whether to put in the half-silvered mirror [beam splitter in
front of detectors] or take it out at the very last minute. Thus one
decides whether the photon ``shall have come by one route, or
by both routes'' after it has ``{\em already done} its travel''.
(quote from \cite{7}).
\end {quote}

With the above in mind, we now proceed to describe
a model for quantum key distribution.
The protocol, which is based on Wheeler's delayed
choice experiment, consists of the following steps:
\\
(1) Alice prepares a sequence of $n$ photons (or low intensity
light pulses), all polarized in one direction. She randomly
inserts the beam splitter $\frac{1}{2} S$ in front of her photons.
In those instances in which she has not inserted the beam splitter
(approximately for $\frac{\displaystyle n}{\displaystyle 2}$ photons),
she randomly sends the photons along route $2a$ or
$2b$. Thus approximately $\frac{\displaystyle n}{\displaystyle 4}$
photon are sent along
route $2a$, and another $\frac{\displaystyle n}{\displaystyle 4}$
photons are sent along
route $2b$.
\\
(2) Bob randomly (and of course, independent of Alice)
inserts his beam splitter $\frac{1}{2}S'$
in front of his detectors.
Thus approximately $\frac{\displaystyle n}{\displaystyle 2}$
photons are detected with the
beam splitter $\frac{1}{2}S'$
in front of the detectors
and another $\frac{\displaystyle n}{\displaystyle 2}$
photons are detected with the
beam splitter removed.
\\
(3) Alice tells Bob
(and to any adversary who may be listening)
in each instance whether she inserted the beam splitter
$\frac{1}{2}S$ in front of her photon.
Bob also announces publicly to Alice in each
instance whether he inserted the beam splitter $\frac{1}{2}S'$ in front
of his detectors.
\\
(4) Alice and Bob discard all instances in which Bob
failed to register a particle.
\\
(5) Bob tests the key distribution by checking that in all instances
in which only one of them (either he or Alice) inserted
the beam splitter (but the other one did not),
detectors $1$ and $2$ were triggered with equal probability,
and in all instances in which they both inserted
their beam splitters,
only detector $1$ was triggered,
i.e., beams $2a$ and $2b$ interfered constructively.
If these conditions are satisfied, then Bob informs Alice that there
was not any eavesdropping. They then keep the data only from
instances in which they both happened to remove their beam splitters
(approximately for $\frac{\displaystyle n}{\displaystyle 4}$ photons).
\\
(6) Alice interprets her data as a binary sequence according to the
following coding scheme:
\\
The photon is sent along route $2a$ = $0$,
\\
The photon is sent along route $2b$ = $1$.
\\
The experimental arrangement is such that when the
beam splitters are removed,
detector $1 (2)$ is triggered, when the photon came by
route $2a (2b)$
(see Fig. $4$ of \cite{7}). Thus
Bob interprets his data as a binary sequence according to the
following coding scheme:
\\
Detector $1$ is triggered, thus the photon came by route $2a$ = $0$,
\\
Detector $2$ is triggered, thus the photon came by route $2b$ = $1$.
\\
(7) With this coding scheme, Alice and Bob have acquired a random
bit sequence with high level of confidence that no one else knows it.

One advantage of the above protocol is that the photons
are all polarized along one axis,
for example along
the $x$ axis. Thus the photons that
Alice sends to Bob (perhaps through a fiber) all suffer the
same transmission loss while in transit (note that transmission
loss depends on polarization). In
contrast, schemes which use photons polarized in different
directions are susceptible to different transmission losses for
different photons.
\pagebreak

\end{document}